\documentclass[twocolumn,english,aps,pra,showpacs]{revtex4}
 \pdfoutput=1
\usepackage[T1]{fontenc}
\usepackage[latin1]{inputenc}
\usepackage{amsmath,wasysym}
\usepackage{graphicx}
\usepackage{amssymb}
\usepackage{esint}

\begin{document}

\title{Angular spin-orbit coupling in cold atoms}


\author{Michael DeMarco\footnote{Current address: Department of Applied Mathematics and Theoretical Physics, Wilberforce Road, University of Cambridge, Cambridge, CB3 0WA, UK}, and Han Pu}
\affiliation{Department of Physics and Astronomy, and Rice Quantum Institute,
Rice University, Houston, TX 77251, USA }


\date{\today}

\begin{abstract}
We propose coupling two internal atomic states using a pair of Raman beams operated in Laguerre-Gaussian laser modes with unequal phase windings. This generates a coupling between the atom's pseudo-spin and its orbital angular momentum. We analyze the single-particle properties of the system using realistic parameters and provide detailed studies of the spin texture of the ground state. Finally, we consider a weakly interacting atomic condensate subject to this angular spin-orbit coupling and show how the inter-atomic interactions modifies the single-particle physics.
\end{abstract}

\pacs{03.75.Mn, 37.10.Vz, 67.85.Bc}

\maketitle

\section{Introduction}
Spin-orbit (SO) coupling has traced a circuitous path through physics. Best known for the $\vec{L}\cdot\vec{S}$ coupling (here $\vec{L}$ and $\vec{S}$ represent the orbital and spin angular momentum of the electron, respectively) that contributes to the atomic fine structure \cite{Foot, Cohen-Tannoudji}, the term was soon applied to the Rashba \cite{Rashba} and Dresselhaus \cite{PhysRev.100.580} coupling between electron spin $\vec{S}$ and its linear momentum $\vec{k}$ present in certain solid state materials. In one of the many recent leaps in ultra-cold physics, Bose-Einstein Condensates (BECs) \cite{Lin:2011aa} and Fermi gases \cite{fermi1, fermi2} have been created with equal parts Rashba and Dresselhaus coupling, and proposals for more varied couplings abound \cite{review1,review2,Galitski:2013aa}. In this paper, we bring SO coupling full circle by introducing a scheme to engineer a coupling between the atomic pseudo-spin and the orbital angular momentum of a cold atom.

The key to creating this angular SO coupling is exchanging the two counter-propagating Gaussian Raman beams used in conventional SO coupling for two co-propagating Raman beams operated in Laguerre-Gaussian (LG) modes. LG beam modes carry orbital angular momentum along the direction of beam propagation \cite{LGBeam}. By choosing beams with unequal phase windings, an orbital angular momentum change may be imparted to atoms transitioning between internal states while the linear momentum change used for conventional SO coupling is annulled by the use of co-propagating beams.

Most excitingly, these systems are within experimental reach. Through the use of holographic techniques or spiral wave plates \cite{oam,Gaunt:2012aa, Pasienski:08}, far-field LG beams can now be created with relative ease. Manipulating cold atoms with LG beams has been studied both experimentally and theoretically, in the context of quantum information storage \cite{store}, slow light propagation \cite{slow}, synthetic gauge fields \cite{gauge}, etc. The experiments that directly motivated our investigations \cite{Vortex1, Vortex2} used LG beams to diabatically write phase windings and spin textures into a BEC, producing coreless vortices and skyrmions in the process. Our predictions are related to these results, though we focus on the adiabatic regime and consider the ground states of these systems. 

Here, we introduce a general formalism for analyzing angular SO coupled cold atoms and present results with low-order LG beams that showcase the unique and interesting properties of these systems. We derive the Hamiltonian and discuss its symmetry properties in Section \ref{sec:ModelGPE}, followed by a discussion of the single-particle spectrum and the properties of the ground state in Section \ref{sec:SingleParticle}. These single-particle studies form the basis for more challenging investigations about the many-body physics. As an example, we present our studies of a weakly-interacting BEC in Section \ref{sec:BEC}. An outlook and concluding remarks are presented in Section \ref{sec:con}.

\section{Model Hamiltonian}\label{sec:ModelGPE}
\begin{figure}
\includegraphics[width=.5\textwidth]{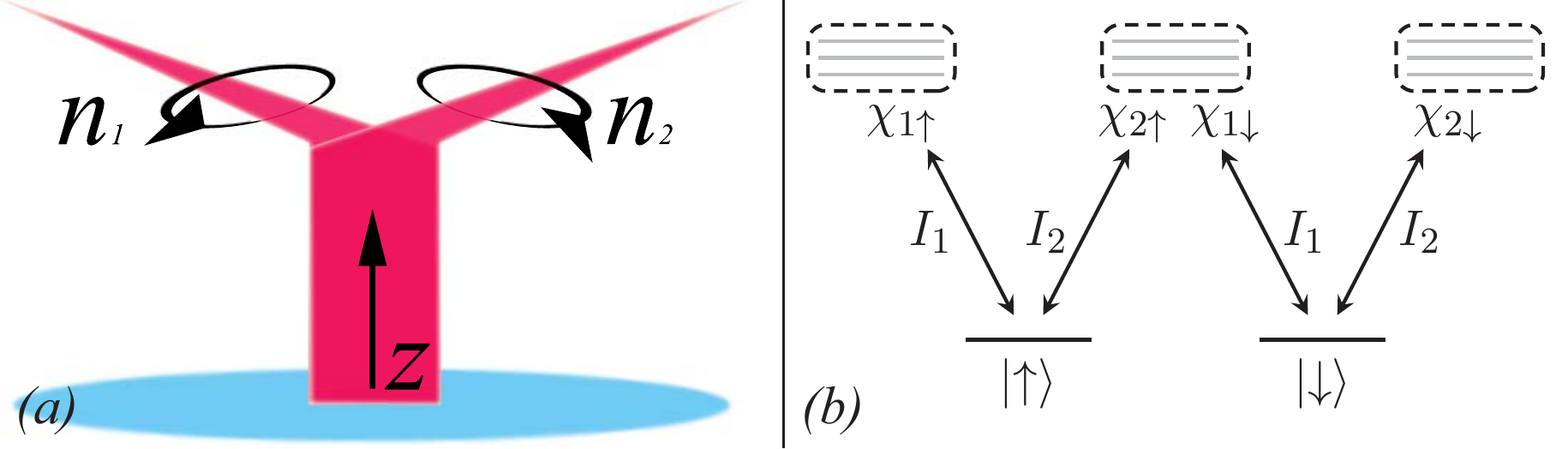}
\caption{(Color online) (a) Schematic representation of the theoretical system. The atomic cloud interacts with two LG beams copropagating in the $-\hat{z}$ direction with phase windings $n_{1}$ and $n_{2}$, intensities $I_1$ and $I_2$, respectively. (b) Two atomic hyperfine ground states, labelled as $|\uparrow \rangle$ and $|\downarrow \rangle$, are coupled by the pair of Raman beams. The beams also induce light shifts with strengths parameterized by the $\chi_{j\sigma}$. \label{fig:schematic}}
\end{figure}

Our theoretical system is schematically shown in Fig.~\ref{fig:schematic}. We consider atoms confined in a two-dimensional (2D) harmonic trap of frequency $\omega$ extending in the $xy$-plane. Two LG beams copropagate in the $-\hat{z}$ direction, coaxial with the center of the trap. LG beam modes \cite{LGBeam} are labelled by two indices $n, m$, and have complex electric field amplitudes at $z=0$ given by: 
\[
{\cal E} (\vec{r})=\sqrt{2I_{0}}e^{-in\phi}\left(\frac{r}{w}\right)^{|n|}L^{|n|}_{m}\left(\frac{2r^{2}}{w^{2}}\right)e^{-r^{2}/w^{2}} \,,
\]
where $L^{|n|}_{m}$ is the associated Laguerre polynomial, $w$ is the width of the beam,  $I_{0}$ describes the intensity of the beam, and we have adopted the cylindrical coordinates $\vec{r}=(r,z,\phi)$. The form for ${\cal E}(\vec{r})$ introduces a non-trivial intensity profile,
\[
\mathcal{I}(r)=I_{0}\left(\frac{r}{w}\right)^{2|n|}\left[L^{|n|}_{m}
\left(\frac{2r^{2}}{w^{2}}\right)e^{-r^{2}/w^{2}}\right]^{2}\,,
\]
while the phase winding $e^{-in\phi}$ reflects the orbital angular momentum $\ell_{z}=-n\hbar$ carried by the beam. 

Via a two-photon Raman process \cite{shore}, the lasers couple two hyperfine states of the atom that we label as $\uparrow, \downarrow$. Under the rotating wave approximation, the following Hamiltonian can be derived \cite{Vortex1, Vortex3, shore}:
\begin{equation}
\hat{H}\Psi=\left[\frac{p^{2}}{2m}+\mathcal{L}(r)+\frac{1}{2}\omega^{2}r^{2}+  \tilde{\Omega}(r)\right]\Psi \,,
\label{eq:nospinrot}
\end{equation}
where
\begin{eqnarray*}
\mathcal{L} &=& \left(\begin{array}{cc} {\cal L}_\uparrow & 0 \\ 0 & {\cal L}_\downarrow \end{array}\right) \\
&=& \left(\begin{array}{cc}\chi_{1\uparrow}\mathcal{I}_{1}+\chi_{2\uparrow}\mathcal{I}_{2} + \delta/2 & 0 \\0 & \chi_{1\downarrow}\mathcal{I}_{1}+\chi_{2\downarrow}\mathcal{I}_{2}-\delta/2 \end{array}\right) \,,
\end{eqnarray*} 
with $\mathcal{I}_{1}(r)$ and $\mathcal{I}_{2}(r)$ being the intensity profiles of the two beams, encodes the light shifts characterized by the coefficients $\chi_{j \sigma}$ ($j=1$, 2 and $\sigma=\uparrow$, $\downarrow$) and also includes the two photon Raman detuning $\delta$, 
\[
\tilde{\Omega}= \Omega(r) \left(\begin{array}{cc}0 & e^{-i(n_1-n_2)\phi} \\e^{i(n_1-n_2)\phi} & 0\end{array}\right)\,,
\]
represents the Raman coupling whose strength is characterized by the parameter $\Omega_0$ with $\Omega(r)=\Omega_{0}\sqrt{\mathcal{I}_{1}\mathcal{I}_{2}}$, and finally $\Psi=(\psi_{\uparrow}, \psi_{\downarrow})^{T}$ is the spinor wave function of the atom.
By measuring mass in units of $m$, energy in units of $\hbar\omega$, and length in units of the oscillator length $\sqrt{\hbar/m\omega}$, the Hamiltonian takes on the dimensionless form:
\begin{equation}
\hat{H}\Psi=\left[-\frac{1}{2}\nabla^{2}+\mathcal{L}(r)+\frac{1}{2}r^{2}+ \tilde{\Omega}(r)\right]\Psi \,,
\label{eq:nodim}
\end{equation}
and we have listed typical values of various parameters in Table~\ref{tab:units}. 

 \begin{table}
 \caption{ Taking $\omega=2\pi \times 1 \text{ kHz}$, the mass of $^{87}$Rb $M=1.443\times10^{-25}\text{ kg}$, and $\hbar$ as the unit measures of frequency, mass, and angular momentum, respectively, fixes the unit measures of length ($a_{{\rm osc}}=\sqrt{\hbar/(m\omega)}$) and energy ($\hbar \omega$). These are presented below in standard units along with typical values of $\chi I$ and $\Omega_{0}I$ in the units used in our calculation. These values are obtained for a typical laser intensity $I=1$ mW/cm$^2$ and a single photon detuning of 1 GHz. \label{tab:units}}
 \begin{ruledtabular}
 \begin{tabular}{c c c c}
 $a_{\rm osc}$ & $\hbar \omega$ &  $|\chi I|$ & $|\Omega_{0} I|$\\ \hline
 $0.34$ $\mu$m & $4.1\times 10^{-12} \text{ eV}$  & $1.8$ & $3.6$\\
 \end{tabular}
 \end{ruledtabular}
 \end{table}

Analogous to the procedure used in conventional SO coupling \cite{Lin:2011aa}, we introduce new basis states $\tilde{\psi}_{\uparrow}=e^{in_{1}\phi}\psi_{\uparrow}$ and $\tilde{\psi}_{\downarrow}=e^{in_{2}\phi}\psi$. Rewriting the Sch\"{o}dinger equation for these states, and with the help of Pauli matrices $\sigma_i$ which act on the atomic pseudo-spin, we have:
\begin{equation}
\tilde{H}\tilde{\Psi}=\left[-\frac{1}{2}\nabla^{2}-\frac{\beta}{r^{2}}\tilde{L}_{z}+\frac{\beta^{2}}{2r^{2}}+\mathcal{L}+
\Omega\sigma_{x}+\frac{1}{2}r^{2}\right]\tilde{\Psi} \,,
\label{eq:spinrot}
\end{equation}
where $\tilde{\Psi}=(\tilde{\psi}_{\uparrow}, \tilde{\psi}_{\downarrow})^{T}$, $\tilde{L}_{z}=-i\partial_{\phi}$ may be regarded as the quasi-angular momentum (QAM) operator, and 
\[
\beta=\left(\begin{array}{cc}n_{1} & 0 \\0 & n_{2}\end{array}\right)=\frac{n_1+n_2}{2} + \frac{n_1-n_2}{2} \sigma_z\,,
\]
is a scaling matrix. Note that the equation is now rotationally invariant, i.e., the QAM operator $\tilde{L}_z$ commutes with the Hamiltonian $\tilde{H}$. Hence each eigenstate of the system possess definite values of QAM characterized by the corresponding quantum number $\tilde{\ell}_{z}$, which is related to the the angular momentum of each spin component in the lab frame by $\ell_{z}^{(\uparrow, \downarrow)}=\tilde{\ell}_{z}-n_{1,2}$. The coupling between atomic pseudo-spin and its orbital angular momentum becomes explicit for $n_1\neq n_2$ as the second term in the square bracket of Eq.~(\ref{eq:spinrot}) contains a term $\propto (n_1-n_2) \sigma_z \tilde{L}_z$. 

The large number of parameters renders a full exploration of the parameter space beyond reach. To focus on the key features of the system, we may constrain a few parameters by consulting current experimental interest (these are tabulated in Table \ref{tab:conventions} for reference). Most experiments using LG beams involve only $n=m=0$ (gaussian) and $|n|=1$, $m=0$ beams, so we focus on the case of $n_{1}=1$, $n_{2}=-1$, and $m_{1}=m_{2}=0$ with red single-photon detuning such that $\chi_{j\sigma}=\Omega_{0}=-1$, and a beam width wider than the oscillator length $w=5$. Lastly, we take the beams to have equal intensity coefficients $I_{10}=I_{20}= I_{0}$ and take the two-photon detuning $\delta=0$.

 \begin{table}
 \caption{In order to reduce the dimensionality of the parameter space to a more manageable size, we fix the values of the following parameters while allowing $I_{10}=I_{20}=I_{0}$ to vary. In the results presented, $\chi_{j\sigma}$, $\Omega_{0}$ and $I$ have been redefined so as to be on the order of $1$ while leaving $\chi I$ and $\Omega_{0}I$ with values on the order of those presented in Table~\ref{tab:units}. \label{tab:conventions}}
 \begin{ruledtabular}
 \begin{tabular}{c c c c c c}
 			\multicolumn{2}{c}{Couplings} & Beam Width & Raman Detuning & \multicolumn{2}{c}{LG indices} \\ \cline{1-2} \cline{3-3} \cline{4-4} \cline{5-6}
			  $\chi_{j\sigma}$ & $\Omega_{0}$ & $w$ & $\delta$ & $n_{1}=-n_2$ & $m_{j}$\\ \hline
			$-1$ & $-1$ & $5 a_{\rm osc}$ & $0$ & $1$  & $0$\\
 \end{tabular}
 \end{ruledtabular}
 \end{table}

As we shall see, the properties of the system are governed by the interplay of the light shifts, the harmonic trap, and the Raman coupling. The light shifts and trap are static potentials (see Figure \ref{fig:disp_rel}), but it is the Raman coupling that enforces the SO coupling. At low intensities, the SO coupling acts as a perturbation on the $I_{0}=0$ simple harmonic oscillator (SHO) ground state, transferring small amounts of population within a given QAM component. For high intensities, the LG light shifts dominate and the condensate forms in a ring centered at $r=0$, which in turn allows for the formation of clouds with higher order phase windings.

\subsection{Symmetries of the Hamiltonian}
If we consider the two spin components to experience the same light shifts (i.e., ${\cal L}_\uparrow = {\cal L}_\downarrow$), then we should expect the system to be invariant under the exchange of the spin labels. This action sends $n\phi\to-n\phi$ as well as inverting the spin space, and is equivalent to reflecting the entire system across the $xy$-plane. For the spin-$1/2$ Hilbert space, this can be represented by the action of $\sigma_{x}$; for the spatial wave function we look for an operator that sends $\ell_{z}\to-\ell_{z}$ while leaving $\vec{r}$ unaffected, that is, a time-reversing (antiunitary) operator. More precisely, we note that the time-reversal operator
\[
\hat{\mathcal{T}}=\sigma_{x}\hat{K} \,,
\]
where $\hat{K}$ denotes complex conjugation, commutes with the Hamiltonian $\hat{H}$ in  Eqs.~(\ref{eq:nospinrot}) and (\ref{eq:nodim}).

This symmetry may be translated into the QAM frame by transforming $\hat{\mathcal{T}}$ under the unitary matrix $U=\text{diag}(e^{in_{1}\phi}, e^{in_{2}\phi})$ to give an operator
\[
\tilde{\mathcal{T}}=U\hat{\mathcal{T}}U^{-1}=e^{i(n_1+n_2)\phi}\sigma_{x}\hat{K}\,,
\] 
that commutes with the Hamiltonian $\tilde{H}$ in Eq.~(\ref{eq:spinrot}).
These symmetries will play an important role in understanding the properties of the ground states in the following sections.

\section{Single-Particle Physics}\label{sec:SingleParticle}

\begin{figure*}
\includegraphics[width=\textwidth]{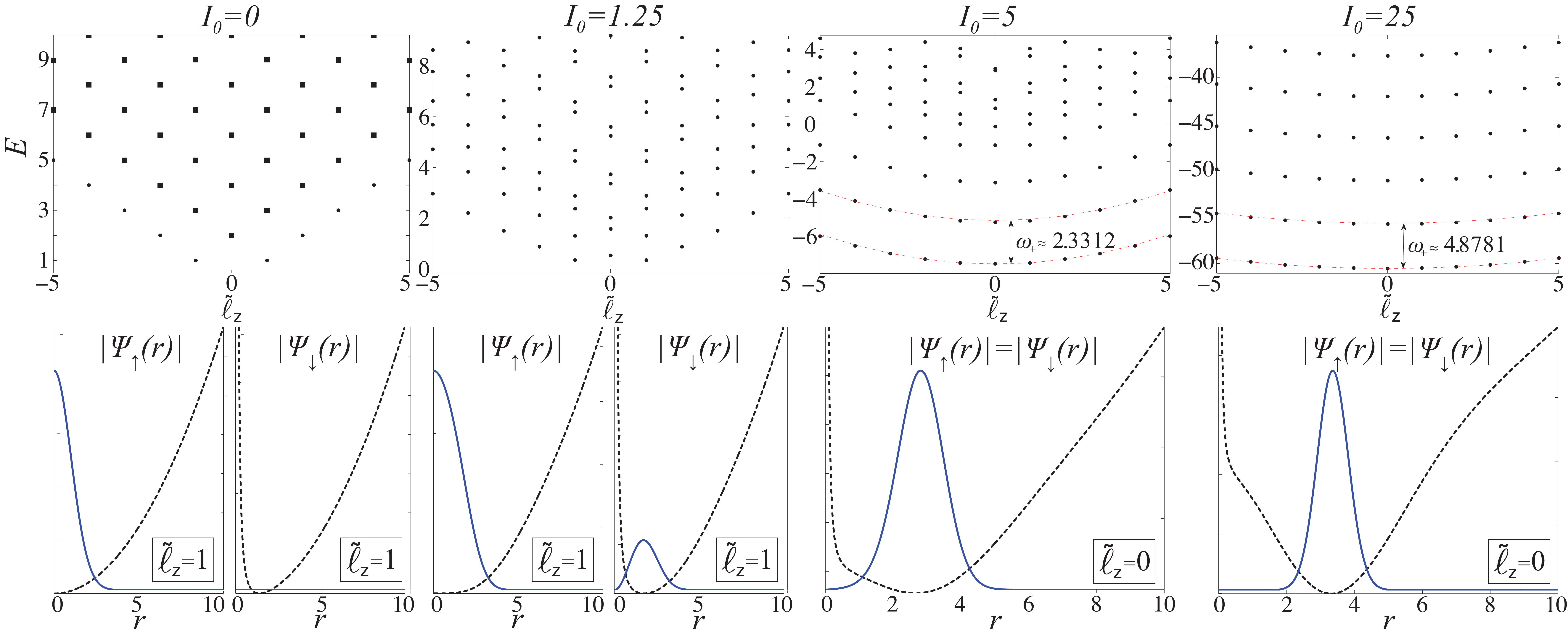}
\caption{(Color online) Upper panel: Single-particle dispersion relations for $I_{0}=0,~1.25,~5,~25$. For the case with $I_0=0$, the dispersion relation is simply that of the 2D Spinor SHO but viewed through the QAM formalism. The doubly degenerate states are shown with black squares. Other states are non-degenerate. The ring-trapped energy predictions of Eq.~(\ref{eq:RingTrapEnergy}) are given by red dotted lines for $I_{0}=5$ and 25. The spectrum is symmetric about $(n_1+n_2)/2=0$. At small $I_0$, the ground state is doubly degenerate with $\tilde{\ell}_z=\pm 1$; at large $I_0$, the ground state is non-degenerate with $\tilde{\ell}_z=0$. The transition occurs at $I_{0}\approx 1.65$. Lower panel: Corresponding effective potentials $V_{\uparrow, \downarrow}(r)$ (black dotted lines) and ground state wave functions (blue solid lines).}\label{fig:disp_rel}
\end{figure*}

By fixing the QAM quantum number $\tilde{\ell}_{z}$, we can mathematically reduce the problem to the radial dimension only. For effective numerical simulation, the divergent terms can be removed by making the ans\"{a}tz
\[
\tilde{\Psi}=e^{i\tilde{\ell}_{z}\phi}\left(\begin{array}{cc}r^{s_{\uparrow}}u_{\uparrow}(r)  \\ r^{s_{\downarrow}}u_{\downarrow}(r)\end{array}\right)\,,
\]
where $s_{\uparrow}=|\tilde{\ell}_{z}-n_{1}|$ and $s_{\downarrow}=|\tilde{\ell}_{z}-n_{2}|$. The single-particle eigenstates may then be determined by applying the finite difference approximation to the equations for $u_{\uparrow, \downarrow}$ and directly diagonalizing the resulting matrix.

We now abandon our general analysis and focus on the specific case of $n_{1}=1$, $n_{2}=-1$, $m_{1}=m_{2}=0$, and all other parameters as in Table \ref{tab:conventions}. The upper panel of Figure~\ref{fig:disp_rel} shows the dispersion relations $E(\tilde{\ell}_{z})$ for this system, and the corresponding ground state wave function is displayed in the lower panel. For the $I_{0}=0$ cases, the band structure present is just that of the spinor simple harmonic oscillator (SHO) viewed in the QAM frame. The ground states found are the expected gaussian wavepackets populating the component with the lab-frame angular momentum $\ell_{z}^{(\uparrow,\downarrow)}=\tilde{\ell}_{z}-n_{1,2}=0$, or QAM $\tilde{\ell}_z = \pm 1$. 

At low light intensity, the twin ground states are perturbed versions of the original $\tilde{\ell}_z=\pm 1$ SHO ground states, with small amounts of population transferred into the previously vacant component. However, at $I_{0}\approx 1.65$, the system transitions to having a single ground state with $\tilde{\ell}_{z}=0$ and $|\psi_\uparrow|=|\psi_\downarrow|$. We will see that this gives rise to a quantum phase transition in the many-particle BEC case.

The symmetry of the bands about $(n_{1}+n_{2})/2$ is guaranteed by the $\tilde{\mathcal{T}}$ symmetry discussed above. In particular, the commutation relation \[ [\tilde{\mathcal{T}}, \tilde{H}]=0 \,,\] together with the anticommutation relation:
\[
\{\tilde{L}_{z}, \tilde{\mathcal{T}}\}=(n_1+n_2)\, \tilde{\mathcal{T}} \,,
\]
determines the effect of $\tilde{\mathcal{T}}$ on the energy eigenstates: If $\tilde{\Psi}$ is an eigenstate of the system with QAM quantum number $\tilde{\ell}_z$, then $\tilde{\cal T} \tilde{\Psi}$ is also an eigenstate with QAM quantum number $n_1+n_2-\tilde{\ell}_z$.
Put simply, $\tilde{\mathcal{T}}$ reflects the spectrum of the system about $\tilde{\ell}_{z}=(n_1+n_2)/2$. Importantly, this implies that any non-degenerate state must have $\tilde{\ell}_{z}=(n_1+n_2)/2$, which is only possible if $(n_1+n_2)$ is even. Hence if we choose, for example, $n_1=1$ and $n_2=0$, then all eigenstates (including the ground state) will remain degenerate.

\subsection{Band Flattening}
From Fig.~\ref{fig:disp_rel}, one can see that as $I_0$ increases, in addition to the ground state changing from two-fold degenerate to non-degenerate, the low-lying dispersion bands becomes flattened. This phenomenon can be attributed to the fact that, for large $I_0$, 
the atoms are confined to a ring-shaped region as a result of the light shifts induced by the red-tuned LG beams. Quantitatively, for a given QAM quantum number $\tilde{\ell}_z$, each spin component is exposed to an effective static potential
\begin{equation}
V_{\uparrow,\downarrow}(r)=\frac{\tilde{\ell}_{z}^{2}}{2r^{2}}+\sigma_{z}\frac{\tilde{\ell}_{z}}{r^{2}}+\frac{1}{2r^{2}}+\mathcal{L}_{\uparrow,\downarrow} + \frac{1}{2}r^{2} \,, \label{veff}
\end{equation}
examples of which are illustrated in the lower panel of Fig.~\ref{fig:disp_rel}.
However, the spatially varying Raman coupling induces further energetic variation across the width of atomic cloud. For $\tilde{\ell}_{z}=0$ and symmetric light shifts with $\mathcal{L}_\uparrow = {\cal L}_\downarrow={\cal L}$, the system may be decoupled by defining the symmetric and anti-symmetric superpositions of the two spin states: $\phi_{\pm}=\frac{1}{\sqrt{2}}(\tilde{\psi}_{\uparrow}\pm\tilde{\psi}_{\downarrow})$, which are governed by their respective Hamiltonian
\[
\tilde{H}_{\pm}=\left[-\frac{1}{2}\nabla^{2}+\frac{1}{2r^{2}}+\mathcal{L}(r)\pm\Omega(r)+\frac{1}{2}r^{2}\right] \,,
\]
with the corresponding effective potentials \[ V_\pm(r) =  \frac{1}{2r^{2}} +{\cal L}(r) \pm \Omega(r) + \frac{1}{2}r^2 \,.\] For our choice of the parameters, $\Omega(r)<0$, hence we will now neglect the $\phi_-$ mode as it has higher energy than the $\phi_+$ mode.
In the limit of high intensity, the effective potential $V_+(r)$ has a deep minimum at radius $r_{\min}\neq 0$, and the atomic density is concentrated in a thin annulus of radius $r_{\min}$. If we consider the ring trap to be infinitely thin, then the eigenenergies will be given by $\tilde{\ell}_z^2/(2r^2_{\rm min})$. A recently work has explored this limit \cite{zhang}. To account for the finite width of the ring, we may expand $V_+(r)$ around $r_{\rm min}$ to second order such that $V_+(r)$ can be approximated as a harmonic potential with freqeuncy $\omega_+$.    Therefore we conjecture that the low-lying eigenenergies can be represented as (apart from a constant shift)
\begin{equation}
E(\tilde{\ell}_{z}, n_{+})=n_+\omega_{+}+\frac{\tilde{\ell}_{z}^{2}}{2r_{\min}^2} \,,
\label{eq:RingTrapEnergy}
\end{equation}
where $n_+=1,2,3,...$ represents the radial quantum number.
The red dotted lines for $I_0=5$ and 25 in the uppper panel of Fig.~\ref{fig:disp_rel} represent the low-lying energy dispersion curves obtained using the above formula, and one can see that they fit with the numerical results extremely well. Note that we obtained $\omega_+$ for states with $\tilde{\ell}_z=0$. For finite $\tilde{\ell}_z$, the effective potential should contain extra contributions from the centrifugal term $\tilde{\ell}_z^2/(2r^2)$. However, as one can see from Fig.~\ref{fig:disp_rel}, this extra term has negligible effect on the value of $\omega_+$.

\begin{figure}
\includegraphics[width=.5\textwidth]{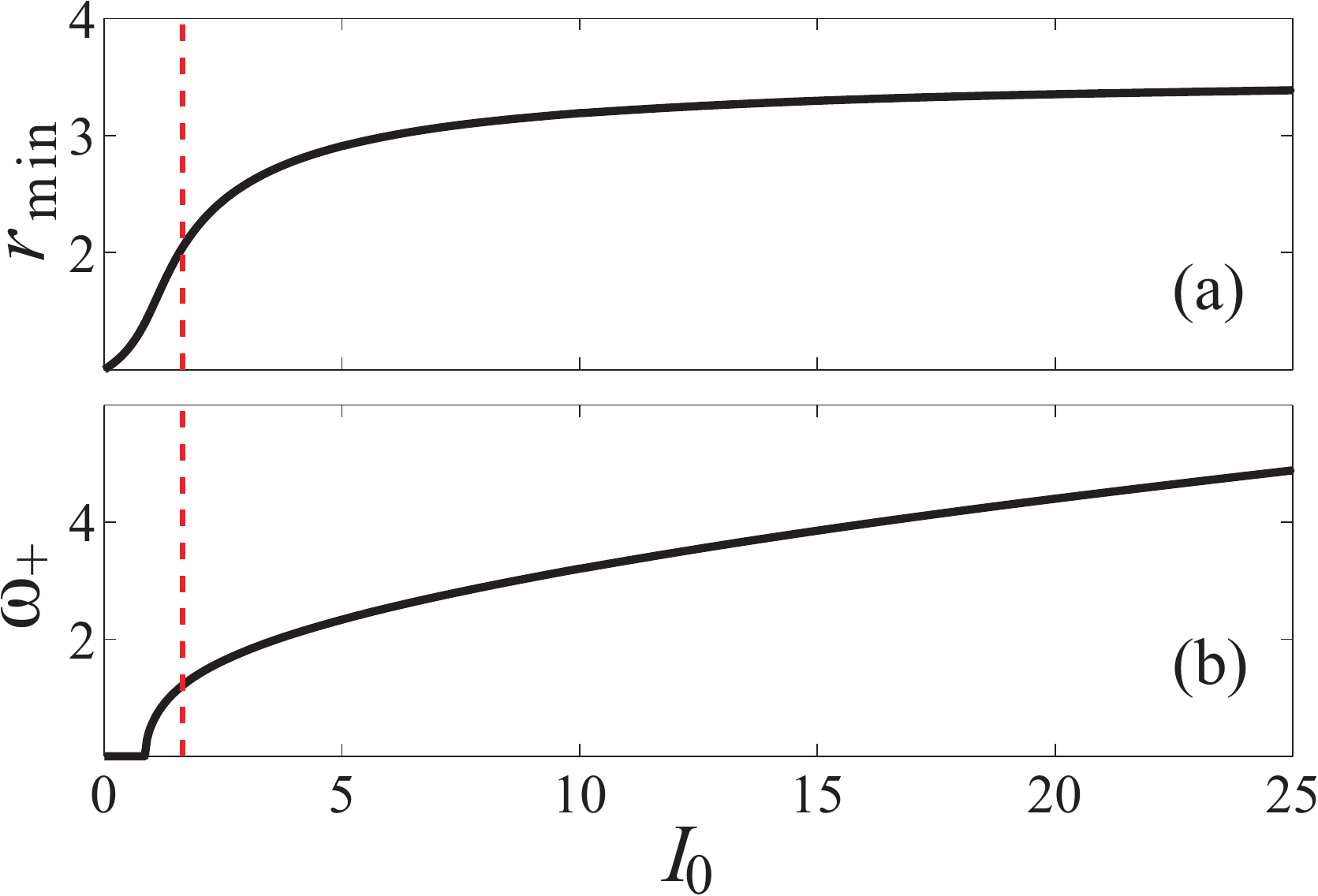}
\caption{(Color online) Dependence of $r_{\rm min}$ (a) and $\omega_+$ (b) on the laser intensity $I_0$. The dashed vertical lines indicate the critical laser intensity at which the ground state changes from two-fold degenerate to non-degenerate.}\label{rmin}
\end{figure}
From Eq.~(\ref{eq:RingTrapEnergy}), we observe that for a given $n_+$, the curvature of the dispersion is determined by the value of $r_{\rm min}$, and the spacing between adjacent bands (with $\Delta n_+=1$) is given by $\omega_+$. We plot in Fig.~\ref{rmin} how these two quantities vary as laser intensity $I_0$ changes. These results show that as $I_0$ increases, $r_{\rm min}$ eventually saturates, whereas $\omega_+$ continues to increase. The infinitely thin ring limit is reached when $\omega_+$ is much larger than all other energy scales of the system.

\subsection{Spin Textures}\label{sec:SpinTex}
The SO coupling gives rise to intriguing spin textures. To characterize the ground state spin texture, we first define a normalized spin vector:
\[
\vec{s}=\frac{\Psi^{\dagger}\vec{\sigma}\Psi}{2|\Psi|^{2}}\,.
\]
Previous studies of 2D Rashba SO coupled \cite{PhysRevA.85.023606} BECs and BECs exposed to LG beams \cite{Vortex2} have found that the spin texture contains a topological knot known as a 2D skyrmion. Obtained from their 3D siblings by stereographic projection, 2D Skyrmions are a subject of interest in BEC studies for the protection that arises from their topological nontriviality. In a skyrmion spin texture, the azimuthal and polar angles of the local spin may be written as $\Theta(r)$ and $\Phi(\phi)$, respectively, which gives rise to the azimuthal $n_{\phi}=(2\pi)^{-1}\Phi(\phi)|_{\phi=0}^{2\pi}$ and radial windings $n_{r}=\cos\Theta(r)|_{r=0}^{\infty}$. The skyrmion number \cite{Sutcliffe} is the topological invariant that distinguishes a skyrmion texture from that of the vacuum; in 2D it is given by:
\begin{equation}\label{eq:skyrmNum}
n_{\text{skyrm}}=\frac{1}{4\pi}\int  \vec{s}\cdot \left(\partial_{x}\vec{s}\times\partial_{y}\vec{s}\right)\,d\vec{r} \,,
\end{equation}
or, in terms of the radial and azimuthal windings \cite{NatureTopologicalSolitons}:
\[
n_{\text{skyrm}}=n_{r}n_{\phi}\,.
\]

\begin{figure}
\includegraphics[width=.5\textwidth]{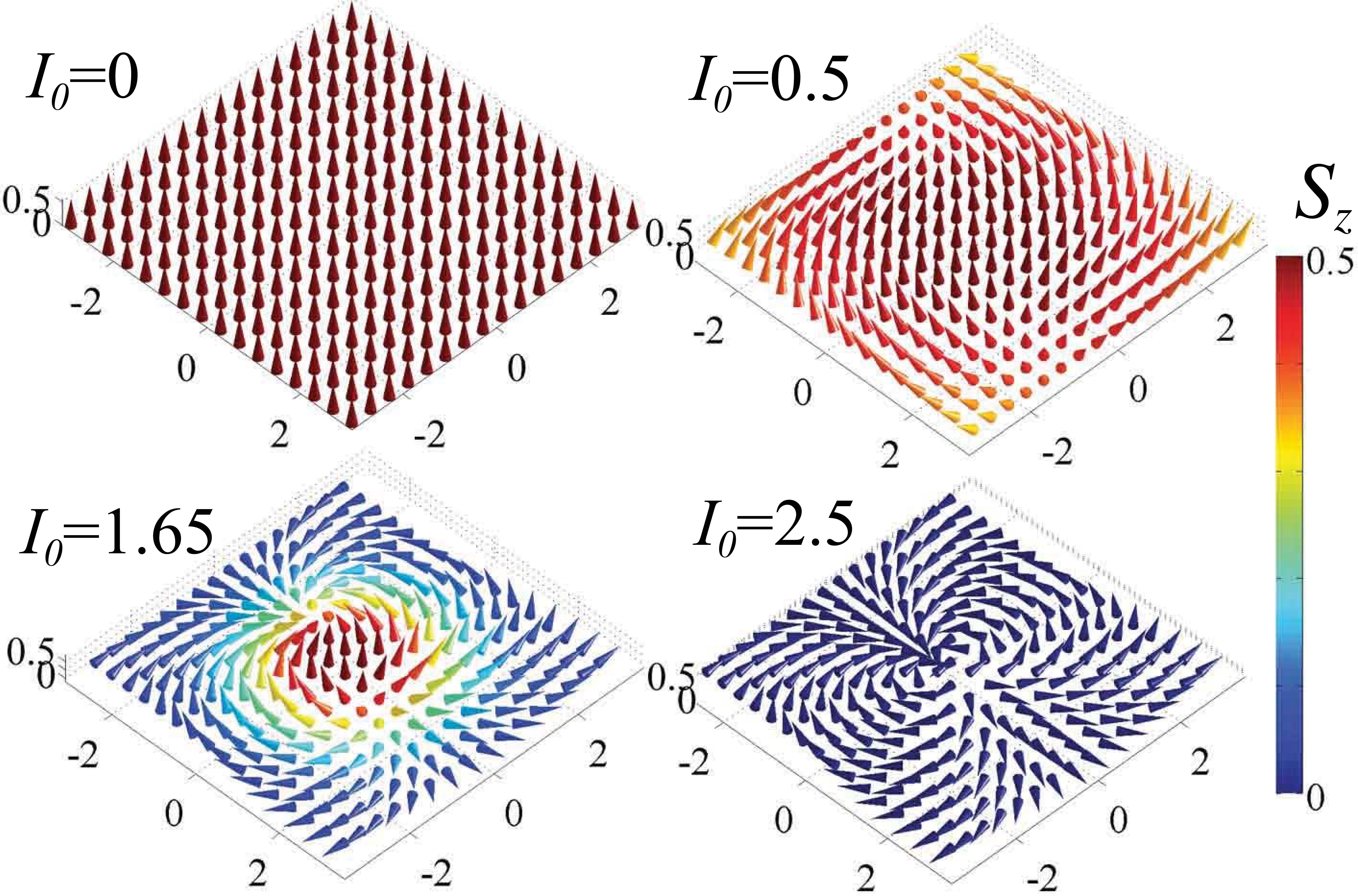}
\caption{(Color online) Spin structure for the ground state. The arrows point in the direction of the local spin $\vec{s}$, and the color represents the spin along the $z$-axis. For $I_0=0$, 0.5 and 1.65, the ground state is degenerate and we pick the one with $\tilde{\ell}_z=1$. For $I_0=2.5$, the ground state is non-degenerate with $\tilde{\ell}_z=0$. The spin structure for $I_0=0.5$ and 1.65 correspond to a half skyrmion with $n_{\phi}=2$ and $n_{r}=1/2$.}\label{fig:SkyrmPic}
\end{figure}

In Fig.~\ref{fig:SkyrmPic}, we present the ground state spin texture at four different values of $I_0$. In our system, for the two-fold degenerate ground states with $\tilde{\ell}_{z}=\pm 1$ at small laser intensity ($I_{0} \apprle 1.65$), we have $n_{\phi}=2$ and $n_{r}=\pm 1/2$, which corresponds to a half skyrmion. As $I_0$ increases from zero, population is transferred into the previously unoccupied component, and the radial winding that forms the skyrmion texture approaches from $r=\infty$, as shown in Fig.~\ref{fig:SkyrmPic}. This manner of formation also bypasses the topological protection usually enjoyed by skyrmions. At large $r$, the atomic density becomes very small and hence, from both a numerical and experimental perspective, $\vec{s}$ is ill-defined. For $I_{0}> 1.65$, the skyrmion spin texture persists in the $\tilde{\ell}_{z}=\pm 1$ components, but the $\tilde{\ell}_{z}=0$ ground state cannot have a skyrmion spin texture, as the $\tilde{\mathcal{T}}$ symmetry implies that $|\psi_{\uparrow}(r)|=|\psi_{\downarrow}(r)|$ and hence the spin becomes planar and lies in the $xy$-plane. Correspondingly, the radial winding $n_r$ and the skyrmion number $n_{\rm skyrm}$ all vanish.

\section{Weakly interacting BEC}\label{sec:BEC}
Our discussion so far has focused on the single-particle physics, which forms the foundation for further exploration of the many-body physics. Here, as a first attempt along this line, we consider a weakly interaction BEC in the mean-field regime.  
An interacting BEC of atoms exposed to the same set-up is described by a Gross-Pitaevksii Equation (GPE) that includes the single-particle Hamiltonian as well as an interaction term:
\[
\mu {\Psi}=\left(\hat{H}+\mathcal{G}\right) {\Psi}\,,
\]
where
\[
\mathcal{G}=\left(\begin{array}{cc}g|\psi_{\uparrow}|^{2}+g_{\uparrow\downarrow}|\psi_{\downarrow}|^{2} & 0 \\0 & g|\psi_{\downarrow}|^{2} + g_{\uparrow\downarrow}|\psi_{\uparrow}|^{2}\end{array}\right)\,,
\]
$g_{\uparrow \downarrow}$ characterizes the inter-species interaction strength,
and we have taken the intra-species interaction strength $g_{\uparrow\uparrow}=g_{\downarrow\downarrow}=g$ for simplicity. To ensure the stability of the condensate, we consider the situation where all interaction strengths are positive.

When we include interactions, the nonlinearity may spontaneously break the rotational symmetry. Hence when solving the GPE, we no longer assume that the system is rotationally symmetric and do the calculation in the 2D $xy$-plane. We determine the ground state by applying a split-step imaginary time evolution \cite{doi:10.1137BAO}, treating the kinetic energy, Raman coupling, and the remaining portions of the GPE in the momentum basis $\psi_{\sigma}(\vec{p})$, QAM basis $\tilde{\psi}_{\sigma}(\vec{r})$, and the usual position basis $\psi_{\sigma}(\vec{r})$, respectively.

\begin{figure}
\includegraphics[width=.5\textwidth]{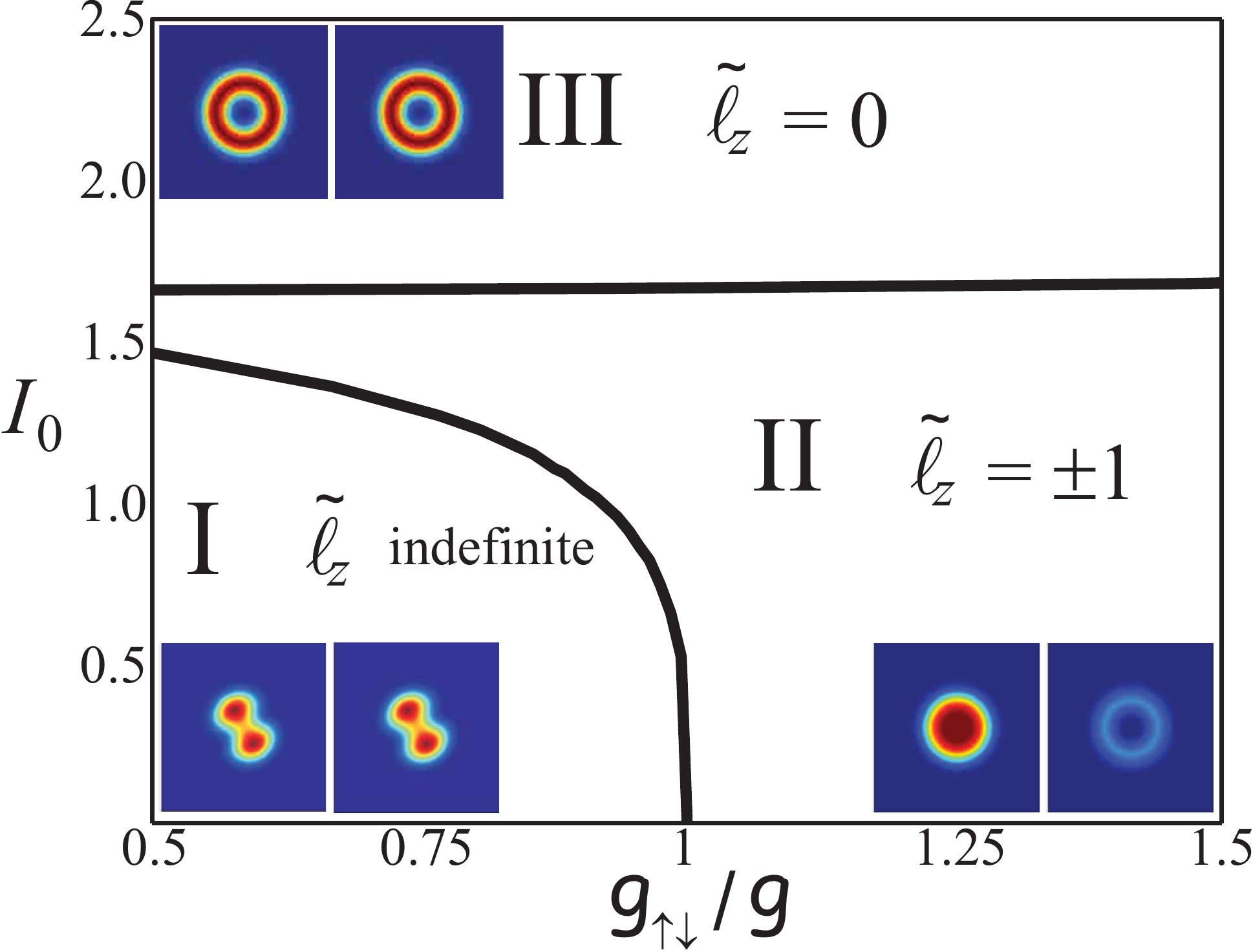}
\caption{(Color online) Phase diagram for the weakly interacting BEC with $g=1$. Insets depict the representative density profiles ($|\psi_\uparrow|^2$ in the left panel and $|\psi_{\downarrow}|^2$ in the right panel) of each phase.\label{fig:phaseDiagram}}
\end{figure}

Figure \ref{fig:phaseDiagram} shows the ground state phase diagram in the parameter space spanned by the laser beam intensity $I_0$ and inter-species interaction strength $g_{\uparrow \downarrow}$ for this system at $g=1$, with insets depicting the representative density profiles of the spin-up and spin-down components in different phases. The three phases found are characterized by their QAM. Phases II and III are the many-body analogs of the single-particle ground states at small and large $I_0$, respectively, and the weak interaction considered here does not change the properties of these states in a qualitative way. Both these phases obey rotational symmetry with definite QAM. The ground state in Phase III is non-degenerate with $\tilde{\ell}_z=0$, while that in Phase II is two-fold degenerate with $\tilde{\ell}_{z}=\pm 1$.

By contrast, as is obvious from the density profiles, Phase I spontaneously breaks the rotational symmetry. Further analysis shows that the Phase I ground state can be regarded as an equal-weight superposition of the two single-particle ground states at $\tilde{\ell}_z=\pm 1$, with an arbitrary relative phase (it can be readily proved analytically that the energy of this equal-weight superposition state is independent of the relative phase). For each realization, this relative phase will be fixed through the mechanism of spontaneous symmetry breaking. In this phase, each spin state can be regarded as a coherent superposition of quantized vortices with different winding numbers \cite{vortex}. Specifically, $\psi_\uparrow$ in the lab frame is a superposition of states with $\ell_z=0$ and $\ell_z=-2$, while $\psi_\downarrow$ is asuperposition of states with $\ell_z=0$ and $\ell_z=2$. However, unlike in previous proposals where such a superposition state is created dynamically \cite{vortex}, here the vortex superposition state represents the ground state of the system.  Furthermore,
for a Phase I state, the density profiles of the two spin components completely overlap with each other, i.e., $|\psi_\uparrow|^2 = |\psi_\downarrow|^2$. Therefore Phase I occurs when $g_{\uparrow \downarrow}$ is small. Increasing the laser intensity creates a more ring-shaped potential which tends to restore the rotational symmetry. This explains the reduced area of Phase I at higher intensities.

\section{Outlook and Conclusion}
\label{sec:con}
In this work, we considered a situation where two hyperfine ground states of an atom are Raman coupled by LG laser beams with different phase windings. This creates a coupling between the atom's pseudo-spin and its orbital angular momentum. Such a situation has already been realized in several experiments, although previous investigations have all focused on the dynamics, instead of the ground state properties that we explored in this work.

We have provided a detailed study of the single-particle physics using realistic parameters. Such studies will form the foundation for the exploration of many-body properties involving a quantum gas. We have performed an investigation of a weakly interacting atomic BEC subject to this angular SO coupling under the mean-field framework. Already in this simple setting, the inter-atomic interactions lead to nontrivial effects. For example, under proper conditions, the interactions spontaneously break the rotational symmetry of the system. Future studies will be extended to stronger interactions which can induce more complicated spin textures \cite{PhysRevA.85.023606,santos} and even lead to strongly-correlated beyond mean-field states \cite{strong}, and also to systems of Fermi gases.

\textit{Acknowledgment} ---This work is supported by the NSF, and the Welch Foundation (Grant No. C-1669). HP acknowledges useful discussions with Nick Bigelow. 

{\em Note added} --- When writing the manuscript, we noticed a preprint by Hu {\em et al.} \cite{half} that considered a similar system as ours. In places where we overlap, our results agree with each other.

\end{document}